\documentclass[16pt,superscriptaddress,showpacs,preprintnumbers,amsmath,amssymb,floatfix,prd]{revtex4}
\usepackage{graphicx}% Include figure files
\usepackage{dcolumn}% Align table columns on decimal point
\usepackage{bm}% bold math
\usepackage{hyperref}
\usepackage{url}
\usepackage{amsfonts}
\usepackage{amsmath}
\usepackage{mathrsfs}
\usepackage{rsfso}
\usepackage{caption}
\usepackage{subcaption}
\hypersetup{colorlinks,citecolor=blue}
\usepackage{float}

\begin{document}

\title{Catenoid Inspired Hyperbolic Wormhole Geometry}

\author{Bikramarka S Choudhury}
\email[Email: ]{ bikramarka@gmail.com}
\affiliation{Department of Mathematics, Jadavpur University, Kolkata 700032, West Bengal, India}

\author{Md Khalid Hossain}
\email[Email: ]{mdkhalidhossain600@gmail.com}
\affiliation{Department of Mathematics, Jadavpur University, Kolkata 700032, West Bengal, India}

\author{Farook Rahaman}
\email[Email: ]{rahaman@associates.iucaa.in}
\affiliation{Department of Mathematics, Jadavpur University, Kolkata 700032, West Bengal, India}

\begin{abstract}
    We unveil a novel class of traversable wormholes exhibiting exact spherical symmetry, geometrically inspired by the minimal surface structure of a catenoid. Introducing the spacetime metric, we rigorously derive its fundamental curvature properties, including the Riemann curvature tensor, and consequently compute the Einstein tensor and stress-energy tensor. This framework reveals that the wormhole is sustained by an anisotropic fluid. A detailed analysis of the energy conditions demonstrates the requisite presence of exotic matter, establishing the physical viability and constraints of this configuration. Subsequent investigations address the wormhole's traversability characteristics, gravitational lensing signatures, and dynamic stability. Crucially, we establish that this catenoid-inspired spacetime represents a finite wormhole, possessing bounded spatial extent.
\end{abstract}

\maketitle

\section{Introduction}

Wormholes \cite{morris-thorne} are currently hypothetical, speculative structures that act as tunnels connecting two distant regions of spacetime. The concept arises naturally within the framework of General Relativity (GR) \cite{GR001}, which revolutionized our understanding of gravity by introducing the notion of curved spacetime geometry. This curvature is governed by the Einstein Field Equations (EFEs) \cite{r1,r2}, which relate the distribution of matter and energy to the geometry of spacetime.

Over the decades, several exact solutions to the EFEs have been extensively investigated. The first non-trivial solution, the Schwarzschild solution \cite{Schwarzschild}, describes the spacetime geometry outside a spherically symmetric non-rotating mass. This solution laid the groundwork for the modern understanding of black holes. Subsequent developments led to the discovery of more general black hole solutions, such as the Kerr \cite{r3} and Reissner–Nordström (RN) metrics \cite{r4}.

In parallel to these black hole solutions, another fascinating class of solutions was explored \cite{black_worm_trans} — those that describe tunnel-like structures in spacetime. Initially introduced as the Einstein–Rosen bridge \cite{PartilcleProblem,r5}, these early models of wormholes were found to be highly unstable. The introduction of exotic matter, proposed in part to account for the observed late-time acceleration of the universe \cite{late_accn}, renewed interest in wormholes. It was later shown that such structures could achieve stability in the presence of exotic matter \cite{exotic01,exotic02,exotic03} that violates the Null Energy Condition (NEC) \cite{evolving01}.

One of the most significant developments in this area was the Morris–Thorne wormhole, which remains a cornerstone of theoretical studies. This model has since been generalized and analyzed under various conditions, including the influence of rotation, cosmological evolution \cite{evolving01}, magnetic fields, gravitational lensing, and traversability criteria. More recently, attention has turned toward quantum wormholes \cite{quantum01}, in attempts to bridge quantum theory and general relativity \cite{r7,r8,quasinormal}.

Recent developments in wormhole physics have introduced a wide range of theoretical models beyond classical general relativity. Traversable wormholes have been explored in nonlinear electrodynamics \cite{c1}, higher-curvature extensions such as Einsteinian cubic gravity \cite{worm_cubic_grav}, Gauss–Bonnet, and 
modified gravities \cite{c3,c5,c6,c11,c12,c40,worm_mod_grav2,worm_f(r)}, as well as hybrid metric–Palatini theories \cite{c4,c7,worm_met_affine}. These models often soften energy condition violations through modified dynamics or exotic matter sources, including string fluids, Casimir energy, and dark matter-inspired profiles in loop quantum cosmology \cite{c8,c10,c13,c30,loop_quantum_gravity}.

Efforts to include more physical phenomenon into the theory of wormholes have led to solutions incorporating rotation \cite{c29,c35}, electric charge \cite{c32}, and topological features such as cosmic strings or monopoles \cite{c2,c28,c34}. These features influence potential observational signals, prompting studies of gravitational lensing, shadow imaging, and light deflection as detection tools \cite{c9,c26,c27,c33,c36}. Meanwhile, insights from quantum gravity—particularly AdS/CFT and holography—have renewed interest in Euclidean wormholes and their role in quantum cosmology \cite{c17,c18,c19}. Novel proposals of traversable wormholes consistent with quantum field theory \cite{c37,c38} and more exotic geometries like ring wormholes or time machines \cite{c16} continue to push the boundaries. Several recent reviews provide a broad overview of these emerging directions \cite{c20,c22}.

In this paper, we propose a novel solution to the Einstein Field Equations: a hyperbolic wormhole geometry inspired by the mathematical structure of a \textit{catenoid}\cite{catenoid}, a classical example of a minimal surface. This geometric influence provides a unique foundation for our wormhole model.

The structure of the paper is as follows. In Section~\ref{sec2}, we present the motivation and formulation of our proposed catenoidal wormhole geometry, initially developed without direct reference to the general wormhole metric. In Section~\ref{sec3}, we compute the Christoffel symbols derived from the metric coefficients. Section~\ref{sec4} contains the calculation of the Ricci tensor and Ricci scalar, which are essential for determining the Einstein tensor and formulating the field equations, presented in Section~\ref{sec5}.

The energy conditions implied by our model are examined in Section~\ref{sec6}. In Section~\ref{sec7}, we provide visualizations of the wormhole geometry for various parameter values, which are kept consistent throughout the paper to maintain coherence—though alternative values may be explored in future work. Section~\ref{sec8} compares our metric with the general Morris–Thorne wormhole metric and verifies the flare-out condition. Traversability criteria are assessed in Section~\ref{sec9} via tidal force calculations. Section~\ref{sec10} explores the deflection of massive particles within the proposed geometry. Finally, in Section~\ref{sec11}, we perform a junction stability analysis of the structure. The paper concludes with Section~\ref{sec12}, which presents a summary and discussion of our findings.

\section{Metric Formulation}\label{sec2}

We propose our metric from the concept of catenoid. Catenoid is a minimal surface obtained from the rotation of the catenary $x_1 = \cosh x_3$ about the $x_3$ axis. The equation of a catenoid embedded in 3D in cylindrical coordinates $(r,\theta,z)$ is
\begin{equation}
    r = \alpha \cosh\left(\frac{z}{\alpha}\right)
\end{equation}

In cartesian coordinates, it can be parametrized as
\begin{eqnarray} \label{cate}
    &x = \alpha \cosh v \cos u, \nonumber \\
    &y = \alpha \cosh v \sin u, \nonumber \\ 
    &z = \alpha v  
\end{eqnarray}
where  $u \in [0, 2\pi]\, \text{ and } v \in (-\infty, \infty)$

Using \eqref{cate}, we get the metric for catenoid as
\begin{equation}
    ds^2 = \cosh^2\left(\frac{v}{\alpha}\right) \left(\alpha^2\,du^2 + dv^2\right).
\end{equation}

\begin{figure}[H]
    \centering
    \includegraphics[width=0.9\linewidth]{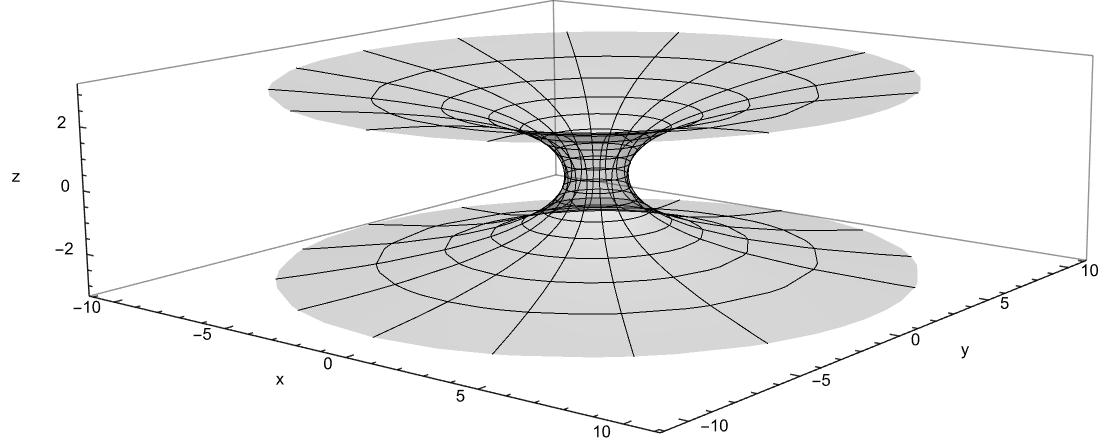}
    \caption{Figure of a catenoid with $\alpha=1$}
    \label{fig:enter-label}
\end{figure}

The metric which is directly derived from catenoidal metric is the Ellis wormhole. If we take the catenoid metric and add spherical symmetry in the other two dimensions, then it would look like:
\begin{equation}
    ds^2 = \cosh^2\left(\frac{r}{\alpha}\right) \left(\alpha^2\,dt^2 + dr^2\right) + r^2(d\theta^2 + \sin^2(\theta)d\phi^2). 
\end{equation}
With a simple transformation in the radial coordinate we can get to the Ellis wormhole which is as follows:
\begin{equation}
    ds^2 = -dt^2 + dr^2 + r^2(d\theta^2 + \sin^2(\theta)d\phi^2).  
\end{equation}
So we think of our construction of metric inspired by catenoid in a slightly different manner by incorporating the idea of Schwarzschild metric. We see, the Schwarzschild metric in geometrical units has the metric coefficients in such a manner that $g_{tt} = -\frac{1}{g_{rr}}$. The Schwarzschild metric is written as:
\begin{equation}
    ds^2 = -\left(1 - \frac{2M}{r}\right) dt^2 + \left(1 - \frac{2M}{r}\right)^{-1} dr^2 + r^2 d\Omega^2.
\end{equation}

Using these two metrics as inspirations, we construct our metric in the following manner with an additional parameter $\beta$:
\begin{equation}\label{metric}
    ds^2 = -\alpha^2 \cosh^2\left(\frac{r}{\alpha}\right) dt^2 + \frac{dr^2}{\cosh^2\left(\frac{r}{\alpha}\right) - \beta} + r^2 d\theta^2 + r^2 \sin^2\theta\, d\phi^2.
\end{equation}

This spacetime has no horizon since the hyperbolic functions are smooth functions. 

Hence the non-zero components of the metric tensor $g_{\mu \nu}$ is defined as:
\begin{equation}\label{g}
    g_{tt} = -\alpha^2 \cosh^2\left(\frac{r}{\alpha}\right)\,,\,\,\, g_{rr}=\frac{1}{\cosh^2\left(\frac{r}{\alpha}\right) - \beta}\,,\,\,\, g_{\theta\theta} = r^2\,,\,\,\, g_{\phi \phi} = r^2 \sin^2\theta
\end{equation}

From this, we calculate $g^{\mu \nu}$. The non-zero components of this contravariant tensor are given as:
\begin{equation}\label{g_inv}
    g^{tt} = -\frac{\text{sech}^2\left(\frac{r}{\alpha }\right)}{\alpha ^2} \,,\,\,\, g^{rr} = \cosh ^2\left(\frac{r}{\alpha }\right) - \beta\,,\,\,\, g^{\theta\theta} = \frac{1}{r^2}\,,\,\,\, g^{\phi \phi}=\frac{\csc ^2(\theta )}{r^2}
\end{equation}

\section{Christoffel Symbols}\label{sec3}

Christoffel symbols are defined as 
\begin{equation}
    \Gamma^\mu_{\nu\lambda} = \frac{1}{2} g^{\mu\rho} (\partial_\nu g_{\rho\lambda} + \partial_\lambda g_{\rho\nu} - \partial_\rho g_{\nu\lambda}).
\end{equation}
We calculate the christoffel symbols by substituting values from  \eqref{g} and \eqref{g_inv} into the above formula. From calculations, the non-zero christoffel symbols are
\begin{eqnarray}
    \Gamma^1_{12} &=& \frac{\tanh \left(\frac{r}{\alpha}\right)}{\alpha}\, ,\,\,\, \Gamma^1_{21}= \frac{\tanh \left(\frac{r}{\alpha}\right)}{\alpha}   \nonumber \\
    \Gamma^2_{11} &=& \alpha  \sinh \left(\frac{r}{\alpha }\right) \cosh \left(\frac{r}{\alpha }\right) \left(\cosh ^2\left(\frac{r}{\alpha }\right)-\beta \right)\, ,\,\,\, \Gamma^2_{22}= -\frac{\sinh \left(\frac{r}{\alpha }\right) \cosh \left(\frac{r}{\alpha }\right)}{\alpha  \left(\cosh ^2\left(\frac{r}{\alpha }\right)-\beta \right)} \, ,\,\,\, \nonumber \\
    \Gamma^2_{33} &=& -r \left(\cosh ^2\left(\frac{r}{\alpha }\right)-\beta \right)\, ,\,\,\, \Gamma^2_{44} = -r \sin ^2(\theta ) \left(\cosh ^2\left(\frac{r}{\alpha }\right)-\beta \right)  \nonumber \\
    \Gamma^3_{23} &=& \frac{1}{r}\, ,\,\,\, \Gamma^3_{32} = \frac{1}{r} \, ,\,\,\, \Gamma^3_{44} = - \sin (\theta ) \cos (\theta ) \nonumber \\
    \Gamma^4_{24} &=& \frac{1}{r}\, ,\,\,\, \Gamma^4_{34} = \cot (\theta) \, ,\,\,\, \Gamma^4_{42} = \frac{1}{r}\, ,\,\,\,\Gamma^4_{43} = \cot (\theta)
\end{eqnarray}

\section{Riemann Curvature Tensor, Ricci Tensor and Ricci Scalar}\label{sec4}

Riemann curvature tensor is defined by the combination of christoffel symbols and derivatives of christoffel symbols
\begin{equation}
    R^\mu_{\nu \lambda \rho} 
= \partial_\lambda \Gamma^\mu_{\nu \rho} 
- \partial_\rho \Gamma^\mu_{\nu \lambda} 
+ \Gamma^\mu_{\sigma \lambda} \,\Gamma^\sigma_{\nu \rho} 
- \Gamma^\mu_{\sigma \rho} \,\Gamma^\sigma_{\nu \lambda}.
\end{equation}

The Ricci tensor is obtained from the Riemann Curvature tensor by contracting its first and third indices:
\begin{equation}
    R_{\mu\nu} = {R^\lambda}_{\mu\lambda\nu}.
\end{equation}

The Ricci scalar is derived from the Ricci tensor through contraction by contravariant metric tensor i.e.
\begin{equation}\label{ricciscalar}
    R= g^{\mu \nu} R_{\mu \nu}
\end{equation}
Where $R$ is the ricci scalar, $R_{\mu \nu}$ is the ricci tensor and $g^{\mu \nu}$ is the contravariant metric tensor.
From this we calculate the Ricci tensor for our metric. The non-zero components of the Ricci tensor are given below:
\begin{eqnarray}
    R_{11} &=& \frac{\cosh \left(\frac{r}{\alpha }\right) \left((r-2 \beta  r) \cosh \left(\frac{r}{\alpha }\right)+2 \alpha  \sinh \left(\frac{r}{\alpha }\right) \left(-2 \beta +\cosh \left(\frac{2 r}{\alpha }\right)+1\right)+r \cosh \left(\frac{3 r}{\alpha }\right)\right)}{2 r} \nonumber \\
    R_{22} &=& -\frac{2 \left(\alpha  \sinh \left(\frac{2 r}{\alpha }\right)+r \cosh \left(\frac{2 r}{\alpha }\right)+\beta  (-r)\right)}{\alpha ^2 r \left(-2 \beta +\cosh \left(\frac{2 r}{\alpha }\right)+1\right)} \nonumber \\
    R_{33} &=& \frac{\alpha  \beta +\alpha +\beta  r \tanh \left(\frac{r}{\alpha }\right)-r \sinh \left(\frac{2 r}{\alpha }\right)-\alpha  \cosh ^2\left(\frac{r}{\alpha }\right)}{\alpha } \nonumber \\
    R_{44} &=& \frac{\sin ^2(\theta ) \left(\alpha  \beta +\alpha +\beta  r \tanh \left(\frac{r}{\alpha }\right)-r \sinh \left(\frac{2 r}{\alpha }\right)-\alpha  \cosh ^2\left(\frac{r}{\alpha }\right)\right)}{\alpha }
\end{eqnarray}

Using these values of Ricci tensor, we get the Ricci scalar by using \eqref{ricciscalar} as:
\begin{equation}
    R=\frac{2 \alpha ^2 \beta +\alpha ^2-\left(\alpha ^2+2 r^2\right) \cosh \left(\frac{2 r}{\alpha }\right)+2 \beta  r^2+4 \alpha  \beta  r \tanh \left(\frac{r}{\alpha }\right)-2 \alpha  r \tanh \left(\frac{r}{\alpha }\right)-2 \alpha  r \sinh \left(\frac{3 r}{\alpha }\right) \text{sech}\left(\frac{r}{\alpha }\right)}{\alpha ^2 r^2}
\end{equation}

\section{Einstein Tensor and Energy-Momentum Tensor}\label{sec5}

The Einstein tensor ($G_{\mu \nu}$) is defined as
\begin{equation}
    G_{\mu\nu} = R_{\mu\nu} - \frac{1}{2} R g_{\mu\nu}
\end{equation}
For the above values of Ricci tensor, Ricci scalar and metric tensor, the non-zero components of Einstein tensor are

\begin{eqnarray}
    G_{11} &=&  \frac{\alpha}{2 r^2}  \cosh ^2\left(\frac{r}{\alpha }\right) \left(2 \alpha  \beta +\alpha -2 r \sinh \left(\frac{2 r}{\alpha }\right)-\alpha  \cosh \left(\frac{2 r}{\alpha }\right)\right) \nonumber \\
    G_{22} &=& -\frac{2 \alpha  \beta +\alpha +4 \beta  r \tanh \left(\frac{r}{\alpha }\right)-\alpha  \cosh \left(\frac{2 r}{\alpha }\right)-r \tanh \left(\frac{r}{\alpha }\right)-r \sinh \left(\frac{3 r}{\alpha }\right) \text{sech}\left(\frac{r}{\alpha }\right)}{\alpha  r^2 \left(-2 \beta +\cosh \left(\frac{2 r}{\alpha }\right)+1\right)} \nonumber \\
    G_{33} &=& \frac{r}{2 \alpha ^2} \left(-2 \alpha  \beta  \tanh \left(\frac{r}{\alpha }\right)+2 r \cosh \left(\frac{2 r}{\alpha }\right)+\alpha  \tanh \left(\frac{r}{\alpha }\right)+\alpha  \sinh \left(\frac{3 r}{\alpha }\right) \text{sech}\left(\frac{r}{\alpha }\right)-2 \beta  r\right) \nonumber \\
    G_{44} &=& \frac{r}{2 \alpha ^2} \sin ^2(\theta ) \left(-2 \alpha  \beta  \tanh \left(\frac{r}{\alpha }\right)+2 r \cosh \left(\frac{2 r}{\alpha }\right)+\alpha  \tanh \left(\frac{r}{\alpha }\right)+\alpha  \sinh \left(\frac{3 r}{\alpha }\right) \text{sech}\left(\frac{r}{\alpha }\right)-2 \beta  r\right)
\end{eqnarray}
The Einstein Field Equations in geometrical units are given by:
\begin{equation}\label{efe}
    G_{\mu \nu} = T_{\mu \nu}
\end{equation}
where $T_{\mu \nu}$ is the energy momentum tensor. 

In this study, we adopt an anisotropic perfect fluid model to describe our system's matter content. Consequently, the energy-momentum tensor is expressed in a diagonal form as
\[
T^\mu_{\nu} = \operatorname{diag}(-\rho,\, p_r,\, p_t,\, p_t),
\]
where \(\rho\) denotes the energy density, \(p_r\) the radial pressure, and \(p_t\) the tangential pressure.
From this we obtain Einstein Field Equations from \eqref{efe}
\begin{equation}\label{eq20}
    \rho =  \frac{1}{2\alpha r^2}  \left(2 \alpha  \beta +\alpha -2 r \sinh \left(\frac{2 r}{\alpha }\right)-\alpha  \cosh \left(\frac{2 r}{\alpha }\right)\right)
\end{equation}
\begin{equation}\label{eq21}
    p_r = -\frac{\left(\cosh ^2\left(\frac{r}{\alpha }\right) - \beta\right).   \left[   2 \alpha  \beta +\alpha +4 \beta  r \tanh \left(\frac{r}{\alpha }\right)-\alpha  \cosh \left(\frac{2 r}{\alpha }\right)-r \tanh \left(\frac{r}{\alpha }\right)-r \sinh \left(\frac{3 r}{\alpha }\right) \text{sech}\left(\frac{r}{\alpha }\right) \right]}{\alpha  r^2 \left(-2 \beta +\cosh \left(\frac{2 r}{\alpha }\right)+1\right)}
\end{equation}
\begin{equation}\label{eq22}
    p_t = \frac{1}{2 \alpha ^2 r} \left(-2 \alpha  \beta  \tanh \left(\frac{r}{\alpha }\right)+2 r \cosh \left(\frac{2 r}{\alpha }\right)+\alpha  \tanh \left(\frac{r}{\alpha }\right)+\alpha  \sinh \left(\frac{3 r}{\alpha }\right) \text{sech}\left(\frac{r}{\alpha }\right)-2 \beta  r\right)
\end{equation}

Up to this point, we have considered a spacetime metric and following the mathematical formalism, we get to the energy density and pressure expressions through the Einstein Field Equations (EFE). Conversely, through the equations \eqref{eq20}$-$\eqref{eq22}, we can say that for such form of matter distribution will lead to a spacetime geometry identical to our proposed metric.

\section{Energy Conditions}\label{sec6}

The \textbf{Null Energy Condition (NEC)} is a fundamental constraint in General Relativity expressed as
\[
T_{\mu\nu} \, k^{\mu} k^{\nu} \geq 0,
\]
for any null vector \(k^{\mu}\) (i.e., \(k^{\mu}k_{\mu} = 0\)). This condition ensures that the energy density, as observed along any light-like trajectory, is non-negative. Essentially, it states that any observer moving along a light-like path should measure a non-negative energy density.

For an anisotropic perfect fluid matter distribution, this condition reduces to 
\begin{equation}
    \rho + p_r \geq 0 \qquad \text{and} \qquad \rho + p_t \geq 0,
\end{equation}
where \(\rho\) is the energy density, \(p_r\) is the radial pressure, and \(p_t\) is the tangential pressure.

\bigskip

The \textbf{Strong Energy Condition (SEC)} is another important criterion in General Relativity. Mathematically, the SEC is expressed as:
\[
\left( T_{\mu\nu} - \frac{1}{2} T g_{\mu\nu} \right) v^{\mu} v^{\nu} \geq 0,
\]
for all timelike vectors \( v^{\mu} \), where \( T = T^{\mu}_{\ \mu} \) is the trace of the energy-momentum tensor.

Through Einstein's field equations, this condition can also be rewritten in terms of the Ricci tensor as:
\[
R_{\mu\nu} v^{\mu} v^{\nu} \geq 0,
\]
which implies that gravity is always attractive, as it leads to the convergence of nearby timelike geodesics.

For an anisotropic fluid configuration, where the radial and tangential pressures \( p_r \) and \( p_t \) may differ, the SEC takes the form:
\begin{equation}
    \rho + p_r \geq 0, \qquad \rho + p_t \geq 0, \qquad \text{and} \qquad \rho + p_r + 2p_t \geq 0,
\end{equation}

\begin{figure}[H]
    \centering
    \includegraphics[width=0.32\linewidth]{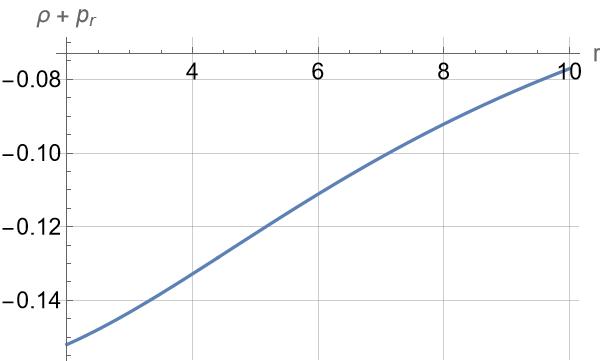}\hfill
    \includegraphics[width=0.32\linewidth]{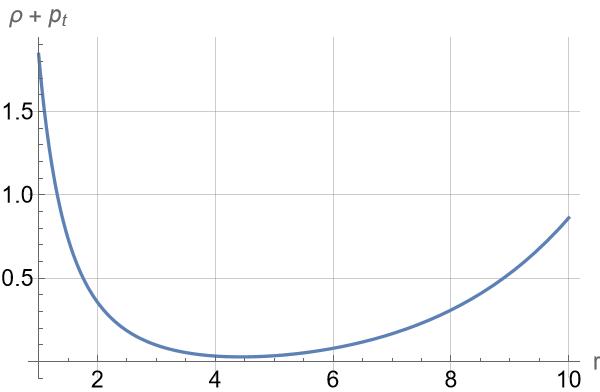}\hfill
    \includegraphics[width=0.32\linewidth]{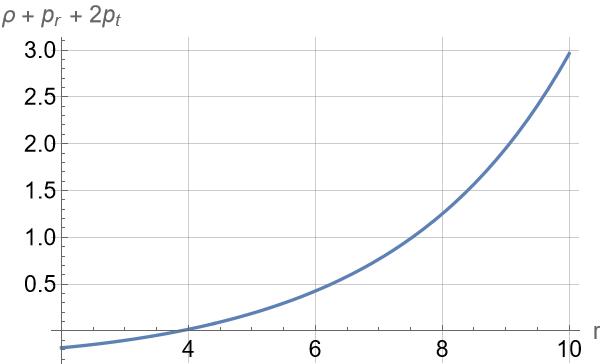}
    \caption{Energy conditions with $\alpha=5$ and $\beta=2$}
    \label{fig:enter-label}
\end{figure}

We see for our wormhole structure, the NEC is violated since $\rho +p_r$ is less than 0 in the region. Also the SEC is violated in a certain region but is satisfied for the rest of the radial values. Overall this indicates presence of exotic matter in the structure which is in alignment with the current theoretical understanding of wormholes.

\section{wormhole visualization}\label{sec7}

From our metric \eqref{metric}, we take \( t = \text{constant} \) and \( \theta = \frac{\pi}{2} \); this reduces the given metric to  

\[
ds^2 = \frac{dr^2}{\cosh^2\left(\frac{r}{\alpha}\right) - \beta} + r^2 d\phi^2.
\]

Now in cylindrical coordinates, the metric is given by

\[
ds^2 = dr^2 + r^2 d\phi^2 + dz^2.
\]

To match the forms, we interpret the given metric as a surface embedded in a 3D space with coordinates \( (r, \phi, z) \). The radial term suggests a transformation for \( dz \):

\[
\frac{dr^2}{\cosh^2\left(\frac{r}{\alpha}\right) - \beta} = dr^2 + dz^2.
\]

This simplifies to

\[
dz = \pm dr \sqrt{\frac{1}{\cosh^2\left(\frac{r}{\alpha}\right) - \beta} - 1}.
\]

The embedding function $z(r)$ is obtained by numerically solving the above equation.

\begin{figure}[H]
    \centering
    \includegraphics[width=0.4\linewidth]{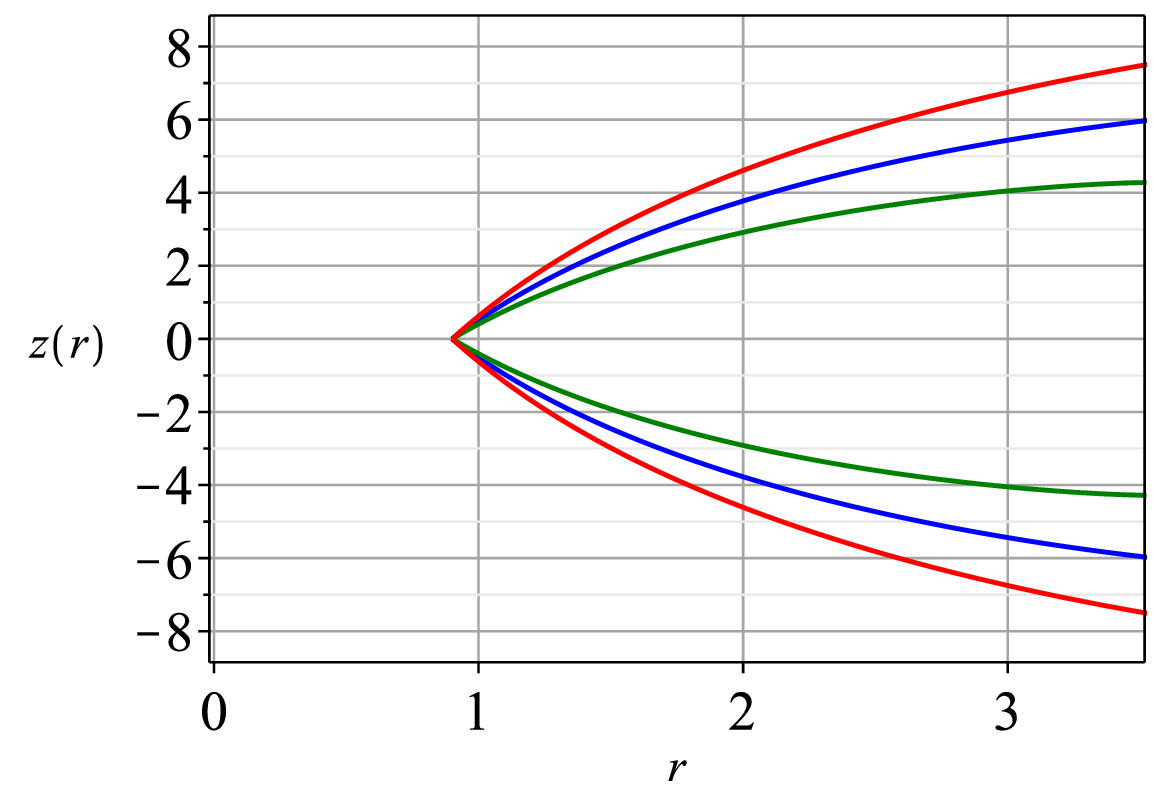}\hfill
    \includegraphics[width=0.6\linewidth]{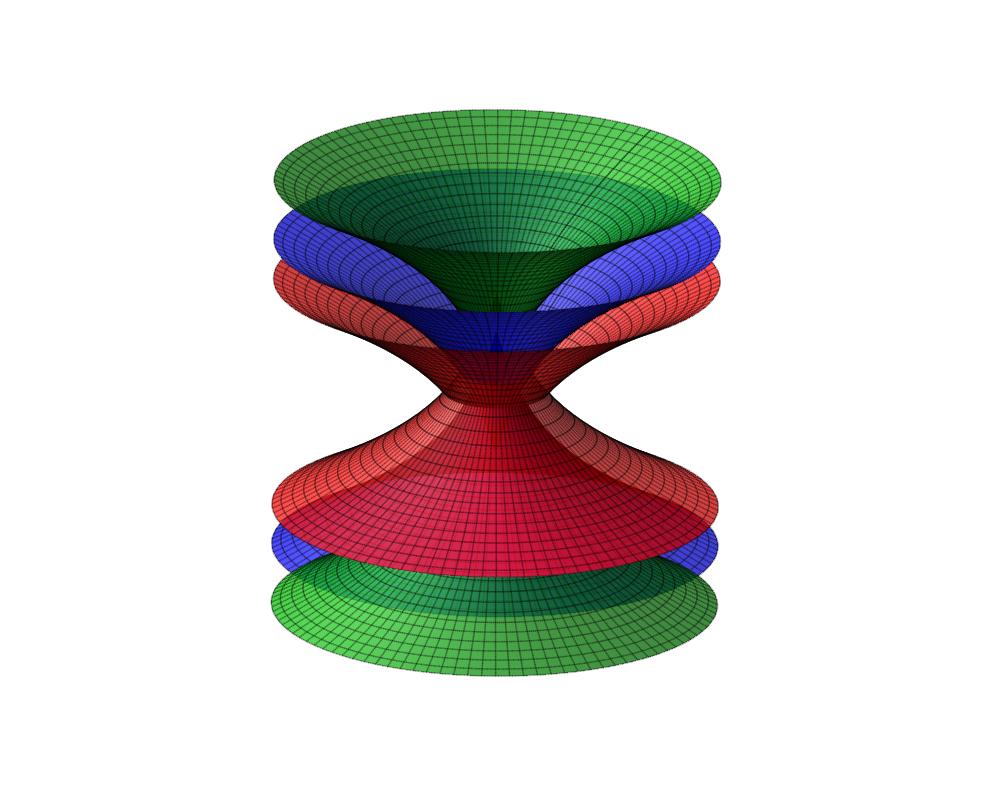}
    \caption{The above diagrams are the graphical representation of embedding surface $z(r)$ (left panel) and visualization of the wormhole (right panel).We assume $\beta=2$ and red for $\alpha=0.8$, blue for $\alpha=1$, and green for $\alpha=1.2$.}
    \label{fig:enter-label}
\end{figure}

\section{Comparison with general wormhole metric}\label{sec8}

\textbf{Derivation from General Wormhole metric:} The metric \eqref{metric} proposed by us can also be derived through the general wormhole metric. We recall the general wormhole metric is given by
\begin{equation}
    ds^2 = -e^{2\Phi(r)} dt^2 + \frac{dr^2}{1 - \frac{b(r)}{r}} + r^2 \left( d\theta^2 + \sin^2\theta \, d\phi^2 \right)
\end{equation}
If we take: $e^{2\Phi(r)} = \alpha^2\cosh^2 \left(  \frac{r}{\alpha} \right)$ and $1-\frac{b(r)}{r} = \cosh^2 \left(  \frac{r}{\alpha} \right) - \beta$, then we can solve for $\Phi(r)$ $\&$ $b(r)$. \\
From above we can write 
\begin{equation}\label{phi}
    \Phi(r) = \frac{1}{2} \ln \left[  \alpha^2\cosh^2 \left(  \frac{r}{\alpha} \right)  \right]
\end{equation}
and
\begin{equation}\label{b}
    b(r) = 1+\beta - \cosh^2 \left( \frac{r}{\alpha} \right)    
\end{equation}
\begin{equation}\label{b'}
   \implies b'(r) = -\frac{1}{\alpha}\sinh\left(\frac{2r}{\alpha}\right)
\end{equation}
Clearly, for $\alpha>0$, the value $b'(r)$ is negative and from the figure below, we see  $|b'| <1$ for $r\leq 5$. It is to  notice that the value of $b'$ doesn't depend on the parameter $\beta$.

\begin{figure}[H]
    \centering
    \includegraphics[width=0.5\linewidth]{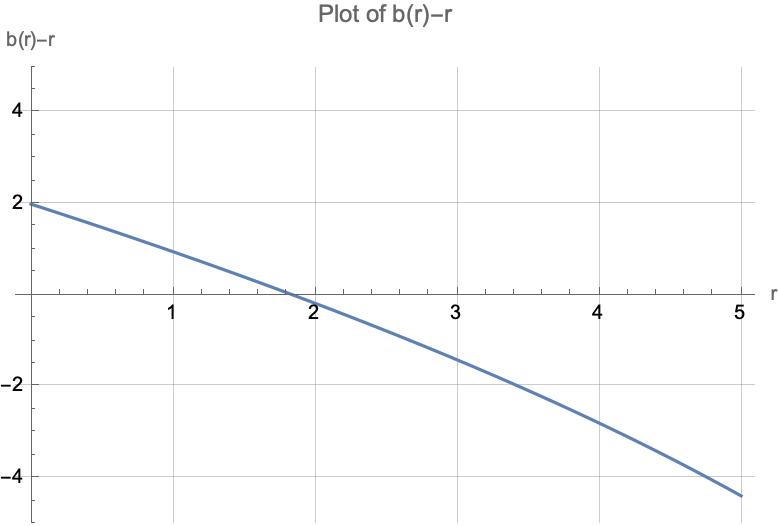}
    \caption{Figure of $b(r)-r$ vs $r$ taking the parameter values as $\alpha=5$, $\beta=2$}
    \label{fig:b(r)}
\end{figure}

\begin{figure}[H]
    \centering
    \includegraphics[width=0.5\linewidth]{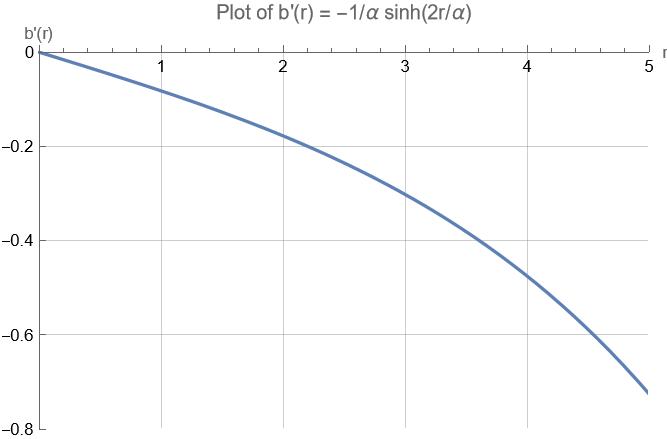}
    \caption{Figure of $b'(r)$ vs $r$ taking the parameter values as $\alpha=5$}
    \label{fig:b'(r)}
\end{figure}

From FIG \eqref{fig:b(r)}, we see $b(r)-r$ cuts the  $r$-axis somewhere between 1.8 and 1.9 (near to 1.87). However finding the precise value is difficult as  the equation $b(r)-r=0$ is a transcendental equation. The point where the plot intersects the $r$-axis is the throat of the wormhole. Also we see that the flare out condition is satisfied as well for a given range of radius. This gives our proposed metric more viability, making it a potentially great topic for study. 

The figure~\ref{fig:b'(r)} indicates that $|b'(r)|<1$. This is sufficient to ensure the flaring-out condition at the throat (where $b(r_0) = r_0$).
In fact, when $|b'(r_0)|<1$, the flaring-out condition is satisfied, meaning that the wormhole geometry opens up.
From the above plot, we see that this criterion is satisfied for $\alpha = 5$ when the radial values are less than $5$.
From this, we conclude that our wormhole is a finite wormhole.\\
At the throat, $r=r_0$, we have
\[
\cosh^2\!\left(\frac{r_0}{\alpha}\right) = \beta,
\]
with $\beta \neq 0$, so the redshift function will never vanish.
Also, since our wormhole is of finite size, the redshift function is finite everywhere, and hence no event horizon is formed.\\
Our constructed model possesses only a finite wormhole interior. Therefore, we do not require global asymptotic flatness.
Instead, one can cut off the wormhole interior at some finite radius $r=a$ and match it to an exterior Schwarzschild solution.
Thus, the wormhole interior is defined only for $r_0 < r < a$.\\
The continuity of the first fundamental form (metric) across $r=a$ requires
\[
e^{2\Phi(a)} = 1 - \frac{2M}{a},
\qquad
1 - \frac{b(a)}{a} = 1 - \frac{2M}{a}.
\]
Hence, the junction conditions imply
\[
b(a) = 2M,
\qquad
\Phi(a) = \tfrac{1}{2} \ln\!\left(1 - \frac{2M}{a}\right).
\]
Thus, the interior wormhole solution is smoothly matched to an exterior Schwarzschild geometry of total mass $M$.
In this construction, the usual requirement of asymptotic flatness is replaced by the above junction condition.

The stability analysis at the junction has been done in the Section \ref{sec11}.

\section{Tidal Forces}\label{sec9}

For a wormhole to be traversable, the tidal forces acting on a body should be well within certain restricted limit. Such calculations has been done by morris-thorne from their paper \cite{morris-thorne}. The paper \cite{evolving01} also has similar calculations.
The radial tidal constraint gives us
\begin{equation}
\left|\left(1 - \frac{b}{r}\right)\left(-\Phi''
+ \frac{b'r - b}{2r(r - b)}\Phi' \;-\; (\Phi')^2 \right)\right|
\;\lesssim\; \frac{g_\oplus}{c^2\, \times \,|\varepsilon| },
\end{equation}

We can take the height of a person traveling through the wormhole to be about 6ft or 2 meter. Here $g_{\oplus}$ is the Earth's gravity. Therefore we take $\varepsilon=2$ meter. Also we have taken $c=1$, so considering $g_{\oplus} \approx 10$, this gives us the RHS of the inequality to be approximately 5.

\begin{equation} \label{radial}
    \implies\Bigg|\frac{-8 r^2-4 \alpha  \beta  \tanh \left(\frac{r}{\alpha }\right)-8 r \cosh \left(\frac{2 r}{\alpha }\right)-3 \alpha  \tanh \left(\frac{r}{\alpha }\right)+\alpha  \sinh \left(\frac{3 r}{\alpha }\right) \text{sech}\left(\frac{r}{\alpha }\right)+8 \beta  r+8 r}{8 \alpha ^2 r^2}\Bigg|\;\lesssim\; 5,
\end{equation}

\begin{figure}[H]
    \centering
    \includegraphics[width=0.5\linewidth]{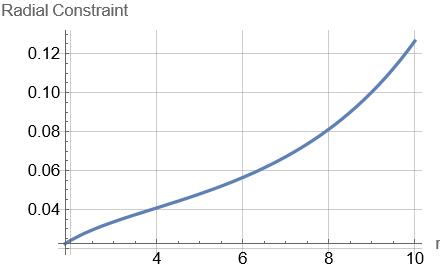}   
    \caption{LHS of the Radial Tidal Constraint for $\alpha=5$, $\beta=2$.}
    \label{fig:radial}
\end{figure}
From the FIG \eqref{fig:radial}, we can evidently see that the radial tidal constraint given by equation \eqref{radial} is satisfied since the values are much lower than $5$ for the values of radius.

And the lateral tidal constraint gives us
\begin{equation}
\left|\frac{1}{\big( 1-\frac{v^2}{c^2} \big)2r^2}\, \left[
 \left(\frac{v}{c}\right)^2
\left( b' - \frac{b}{r} \right) + 2\,(r - b)\,\Phi' \right]  \right|
\;\lesssim\; \frac{g_\oplus}{c^2 \,\times \,|\varepsilon|},
\end{equation}
\begin{equation}\label{lateral}
    \implies\Bigg|\frac{\tanh \left(\frac{r}{\alpha }\right) \left(-2 \beta +\cosh \left(\frac{2 r}{\alpha }\right)+2 r-1\right)-\frac{v^2}{c^2 r} \left(\alpha  \beta +\alpha +r \sinh \left(\frac{2 r}{\alpha }\right)-\alpha  \cosh ^2\left(\frac{r}{\alpha }\right)\right)}{2 r^2 \left(\alpha -\frac{\alpha  v^2}{c^2}\right)}\Bigg|\;\lesssim\; \frac{g_\oplus}{c^2 \,\times \,|\varepsilon|},
\end{equation}

From the lateral tidal constraint, we deduce the inequality for $\frac{v^2}{c^2}$ as:
\begin{equation}\label{velocity}
    \frac{v^2}{c^2} \lesssim \frac{\tanh \left(\frac{r}{\alpha }\right) \left(-2 \beta +\cosh \left(\frac{2 r}{\alpha }\right)+2 r-1\right) + 10r^2 \alpha}{\frac{1}{r}\left(\alpha  \beta +\alpha +r \sinh \left(\frac{2 r}{\alpha }\right)-\alpha  \cosh ^2\left(\frac{r}{\alpha }\right)\right)+10r^2 \alpha}
\end{equation}

\begin{figure}[H]
    \centering
    \includegraphics[width=0.45\linewidth]{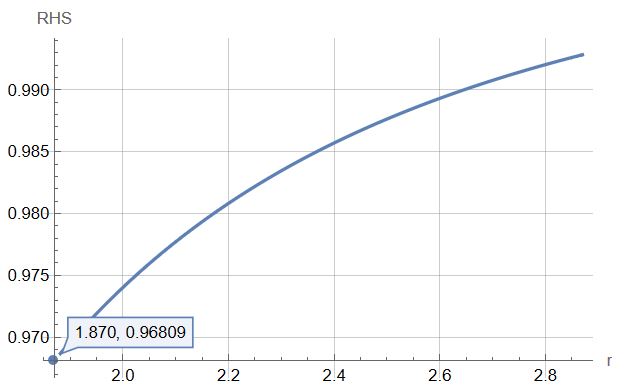}\hfill
    \includegraphics[width=0.45\linewidth]{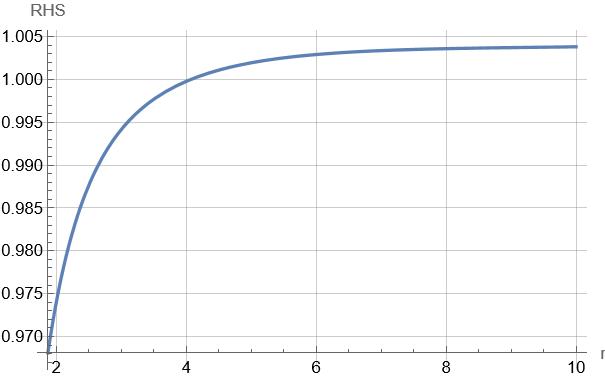}
    \caption{Plot for the RHS of \eqref{velocity} with $\alpha=5$, $\beta=2$. The left image is for radial values close to the throat radius and right image is for radius values up to r=10}
    \label{fig:enter-label}
\end{figure}

From the above figure, we can see that for $\alpha=5$, $\beta=2$, if we keep $\frac{v^2}{c^2} < 0.96$ i.e. $v \lessapprox (0.98)\, c$, then the lateral tidal constraint will be satisfied for all the radial values.

From the above discussion, we see that the wormhole structure satisfies both the radial and lateral tidal constraints which implies that the wormhole is a traversable wormhole.

\section{Deflection of massive particle} \label{sec10}

In this work, we propose a new wormhole metric, and as part of our investigation, we aim to study the deflection of massive particles in this background. To this end, we employ the Rindler-Ishak method, which is particularly well-suited for non-asymptotically flat spacetimes, such as wormholes. This method facilitates the calculation of measurable deflection angles by projecting the particle's path onto the local frame of a static observer. Since the detailed derivations of the geodesic equations and the Rindler-Ishak formalism have already been presented in our previous studies \cite{r10,r11}, we do not repeat them here. Instead, we directly present the trajectory equation governing massive particle motion and derive the corresponding deflection angle in the context of the proposed wormhole geometry.

Trajectory of a massive particle is given by the following expression:
\begin{equation}
U(\phi) = \sin(\phi) \cdot C_2 + \cos(\phi) \cdot C_1 + 
\frac{
    -\cos(\phi) \sin(\phi) \alpha^2 \beta \phi v^2 
    - \alpha^2 \beta \cos^2(\phi) v^2 
    - 2 \cos^2(\phi) b^2 v^2 
    + v^2 \alpha^2 \beta 
    + 2 \cos^2(\phi) b^2 
    + v^2 b^2 
    - b^2
}{2 v^2 b \alpha^2 \sin(\phi)}
\end{equation}

Where, $C_1$ and $C_2$ are arbitrary constants.

\begin{figure}[h]
\centering
\includegraphics[width=7.0cm]{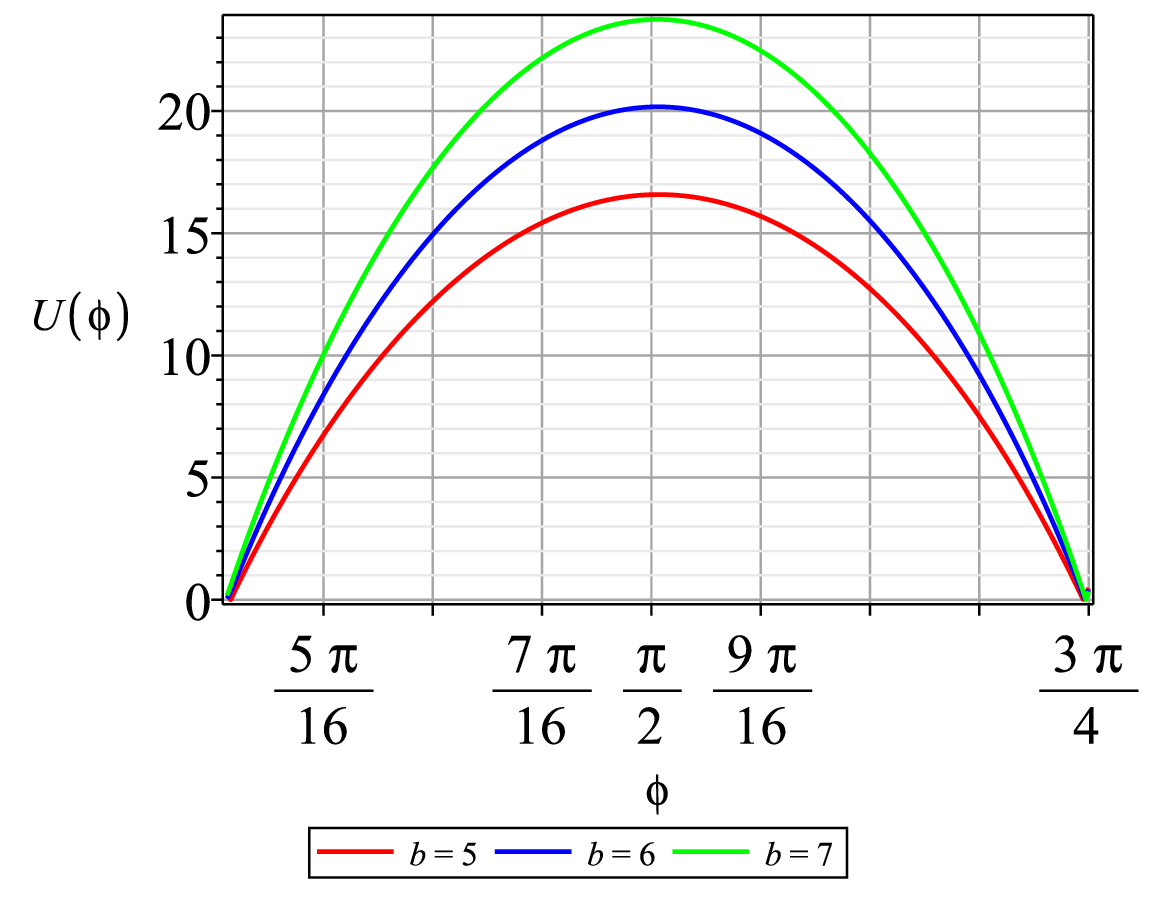}

\caption{The aforementioned diagram show the trajectory $U(\phi)$ as $\phi$ dependent with different values of $b$. In this case, we assume that $\alpha=5$, $\beta=2$, $v=0.7$, $C_1=1$ and $C_2=1$.}
\label{trajectory}
\end{figure}

In Fig.~\ref{trajectory}, we present the trajectories of massive particles around the proposed wormhole, plotted in the $\phi$-coordinate for different values of the impact parameter $b$. It is observed that as $b$ increases, the corresponding particle trajectories extend farther from the wormhole. This behavior signifies that larger values of $b$ correspond to weaker gravitational deflection, leading to trajectories that deviate less and remain at greater distances from the throat of the wormhole.

The deflection angle of a massive particle is given by the following expression:

\begin{equation}
    \alpha^{RI}=2\arctan\left(\frac{\left(\frac{g_{\phi\phi}}{g_{rr}}\right)^{\frac{1}{2}}}{\left|-\frac{1}{U^{2}(\phi)}\frac{dU(\phi)}{d\phi}\right|}\right)-2\phi .
\end{equation}

\begin{figure}[h]
\includegraphics[scale=0.35]{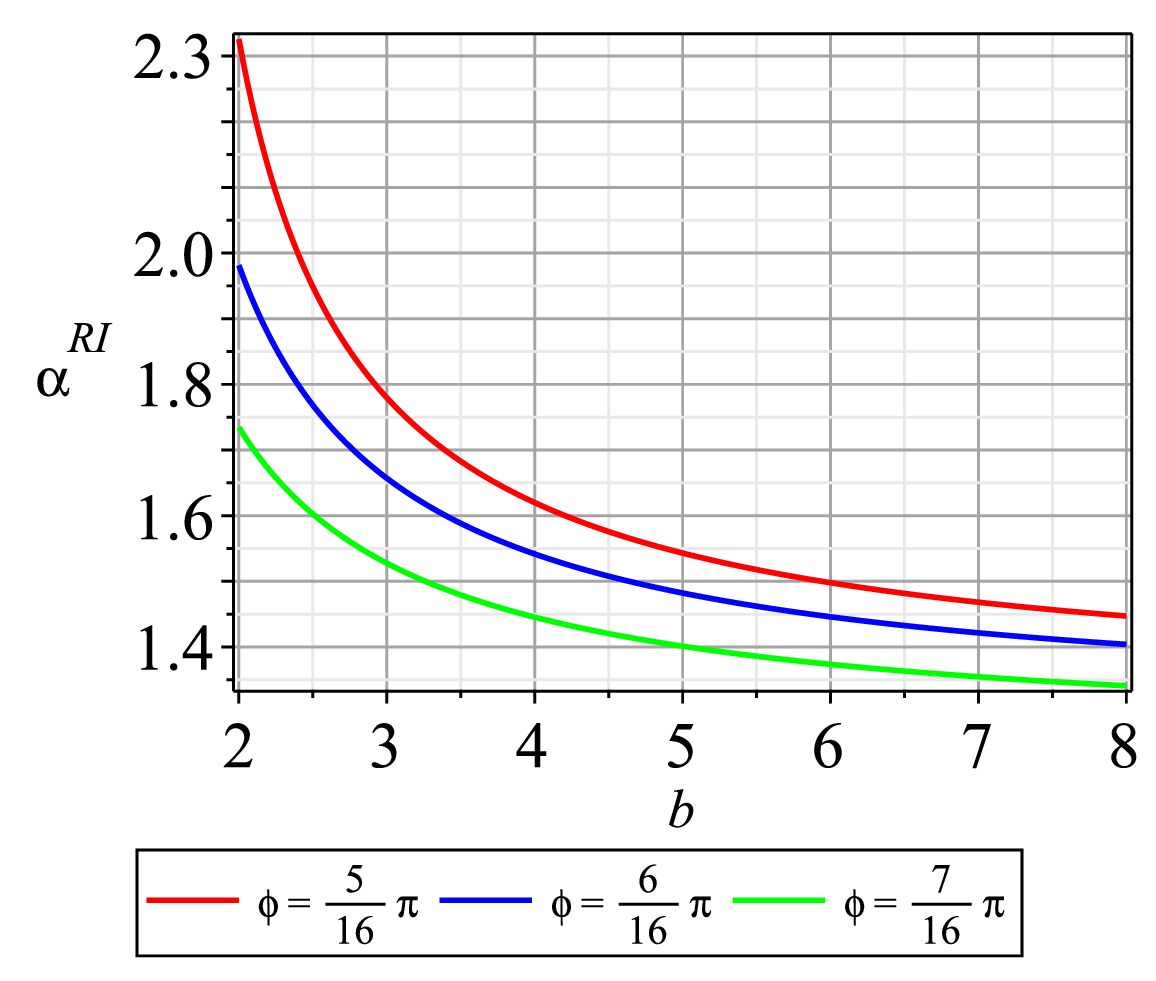}
\includegraphics[scale=0.35]{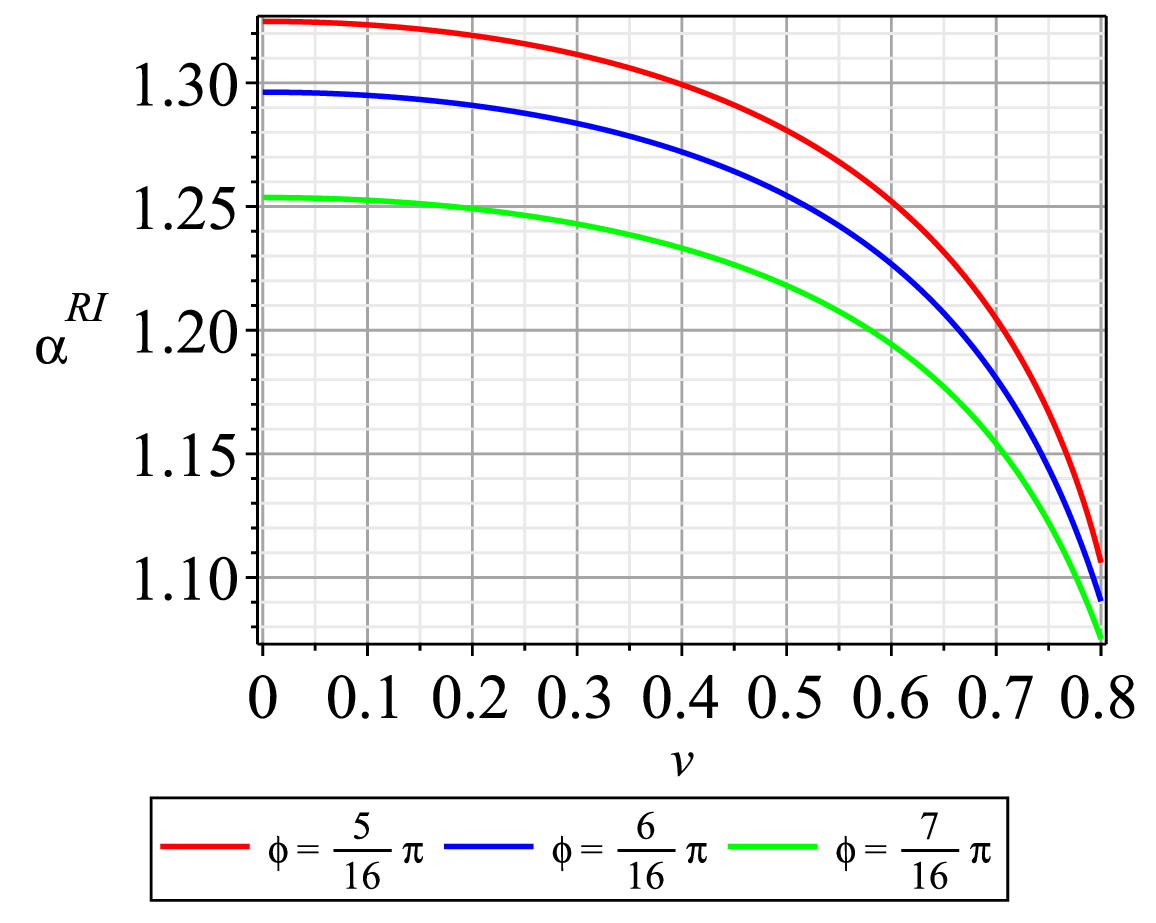}

\caption{Plots for i) Left: Variation of the deflection angle $\alpha^{RI}$ with the impact parameter $b$ for fixed particle velocity $v=0.7$. ii) Right: Variation of the deflection angle $\alpha^{RI}$ with velocity $v$ for fixed impact parameter $b=5$. Here we consider the cases where $\phi$ takes the values $\frac{5}{16}\pi, \frac{6}{16}\pi, \frac{7}{16}\pi$ and the parameter values are $\alpha=5$, $\beta=2$, $C_1=1$ and $C_2=1$.}
\label{deflection}
\end{figure}

From Fig.~\ref{deflection}, we can conclude that the deflection angle for a massive particle decreases as the impact parameter \( b \) increases. However, the rate of this decrease slows down with increasing \( b \). Additionally, the deflection angle also decreases as the particle's velocity increases, but in this case, the rate of decay becomes faster with increasing velocity.

\section{Stability Analysis}\label{sec11}

We take particular values of $\alpha =5$ and $\beta = 2$ for our finite sized wormhole given by \eqref{metric}. We match the metric to the Schwarzschild metric at the boundary. 

We analyze stability through the Darmois–Israel junction formalism \cite{darmois,Israel1966,stability_potential01,stability_potential02}. For this, we have to check the potential function in order to match the two metric functions. Let us assume two separate spacetime manifolds $\mathcal{M}_+$ and $\mathcal{M}_-$ whose metrics are given as:

\begin{equation}\label{metric1}
    d s^{2}=e^{\zeta(r)} d t^{2}-\frac{d r^{2}}{1-\frac{b(r)}{r}}-r^{2}\left(d \theta^{2}+\sin ^{2} \theta d \varphi^{2}\right)
\end{equation}

\begin{equation}\label{metric2}
    d s^{2}=e^{\nu(r)} d t^{2}-\frac{d r^{2}}{1-\frac{B(r)}{r}}-r^{2}\left(d \theta^{2}+\sin ^{2} \theta d \varphi^{2}\right)
\end{equation}

These separate manifolds are bounded by hypersurfaces $\Sigma_+$ and $\Sigma_-$, respectively. A unified manifold $\mathcal{M}$ can be obtained by connecting $\mathcal{M}_+$ and $\mathcal{M}_-$ at their boundaries, resulting in a geodesically complete manifold $\mathcal{M} = \mathcal{M}_+ \cup \mathcal{M}_-$. These two manifolds are connected at the boundaries $\Sigma = \Sigma_+ = \Sigma_-$.

The intrinsic metric on the hypersurface \(\Sigma\) is given by
\begin{equation}
ds^2 = d\tau^2 - a(\tau)^2 d\Omega^2,
\end{equation}

Here $\tau$ is the affine parameter and $a$ is the junction interface.

The Lanczos equation \cite{stability_potential01,stability_potential02} gives the surface stress $\sigma$ as:
\begin{equation}
\sigma = -\frac{1}{4\pi a} \left[ \sqrt{1 - \frac{B(a)}{a} + \dot{a}^2} - \sqrt{1 - \frac{b(a)}{a} + \dot{a}^2} \right],
\end{equation}

The potential function \( V(a) \) is given by \cite{stability_potential01,stability_potential02}:
\begin{equation}\label{potential}
    V(a) = 1 - \frac{b(a)}{a} - \frac{\left(16\pi^{2}\sigma^{2}a^{2} - \dfrac{b(a)}{a} + \dfrac{B(a)}{a}\right)^{2}}{64\pi^{2}\sigma^{2}a^{2}},
\end{equation}

We take \eqref{metric1} to be our catenoidal metric \eqref{metric} and \eqref{metric2} to be Schwarzschild metric. Therefore the quantities of \eqref{metric1} are given by $\zeta(a) = \frac{1}{2} \ln \left[  \alpha^2\cosh^2 \left(  \frac{a}{\alpha} \right)  \right]$ and $b(a) = 1+\beta - \cosh^2 \left( \frac{a}{\alpha} \right)$. Considering the mass of the wormhole to be $M$, we take $\nu(a) = \left(1- \frac{2M}{a}\right)$ and $B(a) = 2M$. We put these quantities in \eqref{potential} to get the potential function for the junction of catenoidal metric with Schwarzschild metric. In the following figure we plot the potential function:

\begin{figure}[H]
    \centering
    \includegraphics[width=0.7\linewidth]{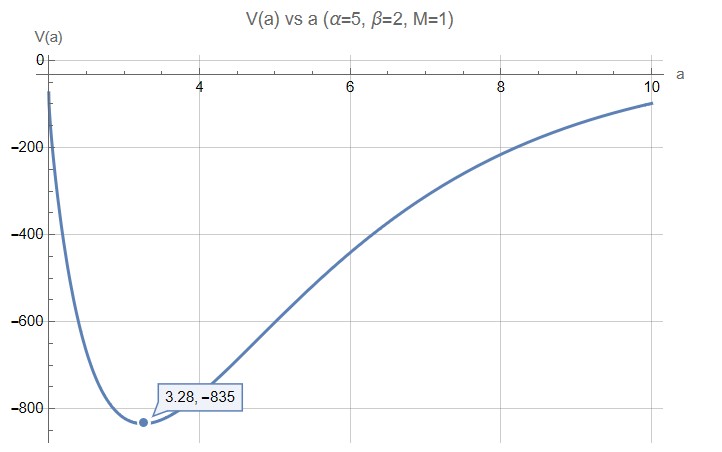}
    \caption{Plot of potential function $V(a)$ given by \eqref{potential} vs junction interface $a$}
    \label{fig:potential}
\end{figure}

From Fig.~\ref{fig:potential} we see that the potential function exhibits a minima for near the junction. Hence our model is stable near to the junction. However, for large value of the junction interface,  the model becomes unstable.

\section{Conclusion and Discussion}\label{sec12}

In this work, we have introduced a novel wormhole geometry motivated by the minimal surface of a catenoid. By introducing this catenoidal profile into four-dimensional spacetime, we derived the corresponding Einstein field equations for a perfect fluid stress–energy tensor and demonstrated that the null energy condition is necessarily violated. This violation underscores the fundamentally exotic nature of our solution. Through an explicit two-dimensional embedding diagram, we visualized the wormhole throat and confirmed that our metric emerges as a special case of the general static, spherically symmetric wormhole line element.

Unlike the canonical Morris–Thorne wormhole, our catenoidal configuration is asymptotically non-flat. We therefore matched its interior to an exterior Schwarzschild geometry, treating the wormhole as a finite-sized bridge between two regions of otherwise vacuum spacetime. We computed the throat radius and verified the flaring-out condition in exact analogy with the standard wormhole construction. Finally, adopting the Morris–Thorne traversability criteria, we evaluated both the radial and lateral tidal accelerations experienced by hypothetical travelers and showed that these remain within acceptable human tolerances. Later we have studied the gravitational lensing characteristic about our wormhole and at the very end, we did a junction stability analysis for the viability of the structure. Thus, our catenoidal wormhole constitutes a geometrically rich yet traversable solution of the Einstein equations.

This study opens several avenues for further investigation. One immediate direction is to explore the stability and dynamical evolution of the catenoidal throat under linear and nonlinear perturbations, both in general relativity and in alternative theories of gravity. The strong gravitational lensing and shadow signatures of this geometry also merit detailed analysis, as they may offer observational discriminants from other exotic compact objects. Moreover, extending the present model to include rotation, charge, or coupling to additional fields could yield a wider class of traversable wormholes with potentially reduced exoticity requirements. We anticipate that these and related studies will deepen our understanding of wormhole physics and its possible phenomenological implications.

\section{Acknowledgment}

BSC, MKH and FR would like to extend their gratitude to JU and IUCAA for their academic support.  FR and BSC also gratefully acknowledge for ﬁnancial support
by SERB, ANRF and UGC, Govt. of India.
We are grateful to the referee for their insightful comments and constructive suggestions, which have significantly enhanced the clarity and quality of this work.

\nocite{*}

\bibliographystyle{unsrt}
\bibliography{refer}

\end{document}